\documentclass{article}
\usepackage{spconf,amsmath,amsfonts,amssymb,graphicx, bm, epstopdf, multicol, subcaption, xcolor}

\def\re#1{\textcolor{black}{#1}}

\title{MACHINE LEARNING PREDICTION OF TIME-VARYING RAYLEIGH CHANNELS}
\name{Joseph Kibugi, Lucas N. Ribeiro, Martin Haardt}
\address{Communications Research Laboratory, Ilmenau University of Technology, Ilmenau, Germany \\ Email: $\{$ joseph.chege, lucas.ribeiro, martin.haardt $\}$@tu-ilmenau.de}
\begin{document}
\ninept
\maketitle
\begin{abstract}
Channel state information (CSI) rapidly becomes outdated in high mobility scenarios, degrading the performance of wireless communication systems. In these cases, time series prediction techniques can be applied to combat the effects of outdated CSI. Recently, it has been shown \re{that} recurrent neural networks (RNNs) exhibit outstanding performance in time series prediction tasks. In this paper, we investigate the performance of RNN and long short term memory (LSTM) predictors in a simple Rayleigh flat-fading channel. We conduct numerical experiments to evaluate whether these machine- learning (ML)-based predictors can outperform the optimal linear minimum mean square error Wiener predictor. Our simulation results indicate that the considered neural network predictors outperform the Wiener predictor for small observation window lengths and are more robust under weak channel correlation \re{as well as} in the presence of noise. Furthermore, we show that simple shallow RNN\re{s} are sufficient to \re{model} Rayleigh channels over a wide \re{range of} Doppler shift\re{s}.
\end{abstract}
\section{Introduction}
\label{sec:intro}

For system data throughput to be maximized, several system parameters need to be adapted to channel conditions. These parameters include, for example, the modulation and coding scheme, transmit power \re{and precoders}. For optimal performance, accurate CSI is required at the transmitter. Acquiring CSI usually involves the transmission of pilot symbols to estimate the channel conditions. However, in high mobility scenarios, the channel undergoes rapid variation and hence the CSI estimated from the pilot symbols quickly becomes outdated, a phenomenon known as channel aging \cite{truongEffectsChannelAging2013}.

A solution to this problem is to predict future channel states based on knowledge of the past channel states. The Wiener predictor \cite{stephenkay} is a well-established prediction method, and it has been shown to compensate for the effects of channel aging in multiple-input multiple-output (MIMO) systems \cite{truongEffectsChannelAging2013}, \cite{zhouHowAccurateChannel2004}. Recently, RNNs and LSTMs have been proposed to solve the channel prediction problem. This is because of their remarkable performance in time series classification and forecasting problems \cite{karimLSTMFullyConvolutional2018}, \cite{liuLSTMBasedClassification2019}. The authors \re{of} \cite{weiliuRecurrentNeuralNetwork2006} investigate narrowband channel prediction using RNNs. In \cite{jiangMultiAntennaFadingChannel2018} and \cite{jiangNeuralNetworkBasedChannel2018}, the authors propose RNN channel predictors in the context of MIMO systems. Comparisons of RNN prediction performance with predictors based on the Kalman filter and autoregressive (AR) models have been made in \cite{jiangLongRangeMIMOChannel2020}, \cite{jiangNeuralNetworkBasedFading2019} and \cite{moralesLinearNonlinearChannel2014} where the RNN predictor provided better prediction accuracy in some scenarios. In \cite{pengLSTMBasedChannelPrediction2020}, an LSTM predictor is used to compensate for the negative effects of imperfect CSI and to ensure secure communications in a massive MIMO vehicular system. The authors in \cite{jiangRecurrentNeuralNetworks2020} leverage LSTM and gated recurrent unit (GRU) architectures to combat outdated CSI in a MIMO system and compare \re{its} performance to that of an RNN predictor.

In this paper, we compare the channel prediction performance of the classical Wiener filter and modern ML solutions through computer simulations. \re{In particular,} we determine experimentally \re{under which conditions} the considered ML-based predictors can outperform the Wiener predictor, which is optimal in the mean square error (MSE) sense. In so doing, we seek to assess the common claim that ML predictors generally outperform the classical model-based solutions. We also investigate how deep a neural network needs to be for the considered problem. Our results indicate that shallow RNNs sufficiently \re{model} the channel over a wide \re{range of} Doppler shift\re{s}. Furthermore, we observe that the RNN and LSTM predictors can provide more accurate prediction \re{results} than the Wiener filter with only a few observed samples. For instance, for a prediction length of 1 ms and observation window lengths smaller than 5 samples, the RNN predictor has a MSE gain of about 1 dB over the Wiener predictor. The ML-based predictors also perform better under weak channel correlation. For example, the RNN predictor exhibits a performance gain for prediction lengths greater than 3 ms with an observation window of 5 samples. The ML-based predictors are more robust in the presence of noise, providing a performance gain even at low signal-to-noise ratios (SNRs). \re{Nevertheless, the considered methods perform roughly the same for some scenarios, e.g., larger prediction lengths with only a few observed samples.}

\textit{Notation:} Vectors and matrices are written as lowercase and uppercase boldface letters, respectively. The transpose and the Euclidean norm of a vector are represented by $[\cdot]^T$ and $\|\cdot\|$, respectively. Moreover, $\mathrm{j}$, $\mathbb{C}$, $\mathbb{R}$ and $\mathbb{E}$ represent the imaginary unit, the set of complex numbers, the set of real numbers, and the expectation operator, respectively.

\vspace{-0.25cm}

\section{System Model}
\label{sec:chanmod}

We consider a wireless system in which a single-antenna access point (AP) communicates with a single-antenna user equipment (UE). The system operates on perfectly synchronized time-division duplexing (TDD). Hence, the base station (BS) can estimate the channel samples from uplink reference signals and use them to predict the future states of the downlink channel. The symbol period is $T_s$ and the system bandwidth is $B=1/T_s$. Assuming a time-varying flat-fading channel, the discrete-time representation of the \re{baseband signal received} by the AP at symbol period $n$ is
\begin{equation}
    y[n]=h[n]x[n]+w[n],
\end{equation}
with $h[n]$ representing the channel coefficient, $x[n]$ the transmitted symbol, and $w[n]$ the zero-mean circularly symmetric complex Gaussian (ZMCSCG) noise term with variance $\sigma_w^2$, respectively. 

We assume that the UE transmits orthogonal pilot symbols $x[n]$, $\mathbb{E}\{|x[n]|^2\}=1$, and that the AP applies least-squares (LS) estimation to obtain the uplink channel coefficients. The LS estimate can be expressed as
\begin{equation}
 \tilde{h}[n] = y[n]x^{-1}[n] = h[n] + w[n]x^{-1}[n] = h[n] + \tilde{w}[n].  
\end{equation}
Note that the variance of $\tilde{w}[n]=w[n]x^{-1}[n]$ is the same as that of $w[n]$, i.e., $\sigma_w^2$, since $x[n]$ has unit power. The AP stores $L$ past noisy LS estimates in $\tilde{\bm{h}}\re{[n]}=\big[\tilde{h}[n],\dots,\tilde{h}[n-L+1]\big]^T \in \mathbb{C}^L$ to predict the channel state $\ell$ symbol periods ahead, i.e., $\hat{h}[n+\ell]$. We define the SNR as 
\begin{equation}
    \mathrm{SNR} = \frac{\|\bm{h}\re{[n]}\|^2}{\sigma_w^2},
\end{equation}
where $\bm{h}\re{[n]}=\big[h[n],\dots,h[n-L+1]\big]^T \in \mathbb{C}^L$ is the noiseless channel vector over the past $L$ symbol periods.

\begin{figure}[t]
    \centering
    \includegraphics[width=0.9\columnwidth]{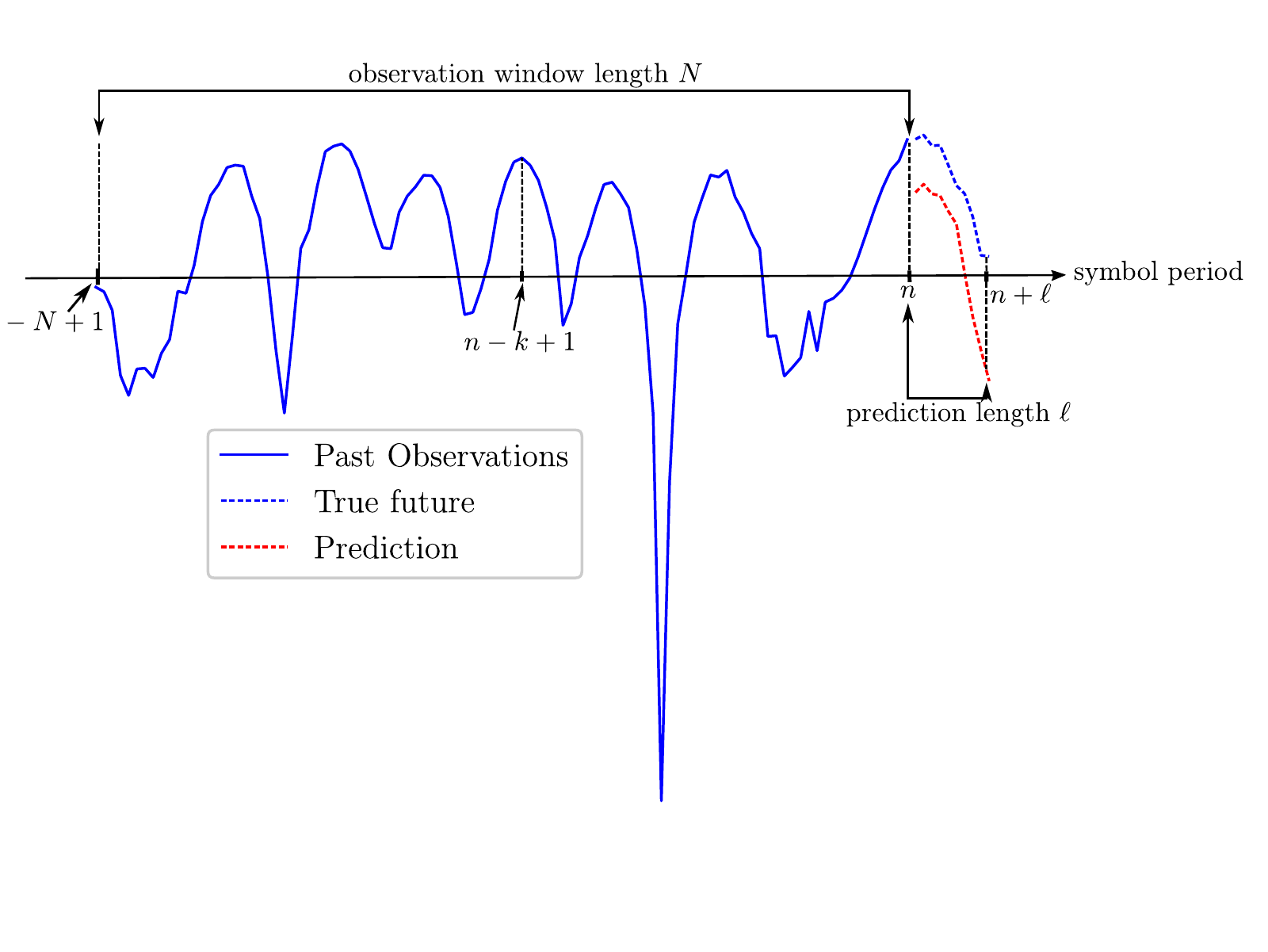}
    \caption{Illustration of channel prediction.}
    \label{fig:predchan}
\end{figure}

The $L$ estimates are divided into two parts, i.e., $L_{\mathrm{train}}$ estimates \re{are} used to calculate the autocorrelation function and to train the neural networks, and $L_{\mathrm{test}}$ estimates \re{are used to test} the predictors. From $L_{\mathrm{test}}$, $N$ estimates, referred to as the predictor order or the observation window length, are used to predict a future channel sample. These details are further explained in Section \ref{sec:chanpred}. Figure \ref{fig:predchan} illustrates the parameters discussed so far.

\subsection{Channel Model}
The sum-of-sinusoids (SOS) method of \cite{popLimitationsSumofsinusoidsFading2001} is used to simulate correlated wide-sense stationary (WSS) Rayleigh fading channels. \re{The} SOS \re{method} represents the channel as a superposition of randomly-phased sinusoids. Hence, $h[n]$ can be written as
\begin{align}
   h[n] \triangleq h(nT_s) = \frac{1}{\sqrt{M}}\sum_{m=1}^{M}\left[x_{\mathrm{I},m}(nT_s)+
   \mathrm{j}x_{\mathrm{Q},m}(nT_s)\right],
   \label{eqn_sos}
\end{align}
where $M$, $x_{\mathrm{I},m}(nT_s)$ and $x_{\mathrm{Q},m}(nT_s)$ represent the number of sinusoids, the $m$-th in-phase component and the $m$-th quadrature component, respectively. Further, $x_{\mathrm{I},m}(nT_s)$ and $x_{\mathrm{Q},m}(nT_s)$ are given by
\begin{subequations}
\small
\begin{align}
   x_{\mathrm{I},m}(nT_s) = A_m\cos{[(2\pi f_{\mathrm{D,max}}nT_s+\psi_m)\cos{(\alpha_m)}+\phi_m]} \label{eqn_sos_1}, \\
   x_{\mathrm{Q},m}(nT_s) = B_m\sin{[(2\pi f_{\mathrm{D,max}}nT_s+\psi_m)\cos{(\alpha_m)}+\phi_m]},
   \label{eqn_sos_2}
\end{align}
\end{subequations}
with $A_m$ and $B_m$ representing Gaussian-distributed random attenuations with zero mean and unit variance, $\alpha_m$ and $\phi_m$ the angle of arrival and the phase shift for the $m$-th path, and $f_{\mathrm{D,max}}$ the maximum Doppler shift in Hertz. To ensure that the simulator produces WSS random variables, an additional phase term $\psi_m$ is required~\cite{popLimitationsSumofsinusoidsFading2001}. The terms $\alpha_m$, $\phi_m$ and $\psi_m$ are uniformly distributed within the interval $[-\pi,\pi)$. It is known that the SOS model (\ref{eqn_sos}) approaches the well-known Jakes model as the number of sinusoids $M$ increases~\cite{popLimitationsSumofsinusoidsFading2001}. The discrete-time autocorrelation function for the Jakes model is given by
\begin{align}
   \varphi_{h\re{h}}[k]=\mathbb{E}\{h[n]h^{\ast}[n+k]\}=J_o(2\pi f_{\mathrm{D,max}}|k|T_s), 
   \label{acf}
\end{align}
and the Doppler power spectrum by
\begin{align}
    \Phi_H(f_D)=\frac{1}{\pi f_{\mathrm{D,max}} \sqrt{1-\left(\frac{f_{\mathrm{D}}}{f_{\mathrm{D,max}}}\right)^2}}, 
\end{align}
where $J_o(\cdot)$ represents the zeroth-order Bessel function of the first kind, and $k$ the lag \re{index.}

\section{Channel Prediction}
\label{sec:chanpred}

\subsection{Wiener Predictor}
The $N$-th order Wiener predictor consists of a linear filter that combines $N$ past LS estimates to predict $\ell$ steps ahead. The estimated channel sample is given \re{by}
\begin{align}
    \hat{h}[n+\ell]=\sum_{k=1}^{N}w^{\ast}_{k,\ell}\tilde{h}[n-k+1],
    \label{wiener}
\end{align}
where $\{w_{1,\ell},\ldots,w_{N,\ell}\}$ are the predictor coefficients. The Wiener predictor minimizes the MSE, defined as
\begin{equation} \label{eq:mse}
    \mathrm{MSE}=\mathbb{E}\big\{\big|h[n+\ell]-\hat{h}[n+\ell]\big|^2\big\}.
\end{equation}
The predictor coefficients $\bm{w}_{\ell}=[w_{1,\ell},w_{2,\ell},\dots,w_{N,\ell}]^T$ which minimize \eqref{eq:mse} are given by
\begin{align}
    \bm{w}_{\ell}=\bm{R}_{\tilde{h}\tilde{h},\ell}^{-1}\bm{r}_{\tilde{h}\tilde{h},\ell},
    \label{pred_eqn}
\end{align}
in which $\bm{R}_{\tilde{h}\tilde{h},\ell} \in \mathbb{C}^{N \times N}$ denotes the correlation matrix of the LS samples, and  $\bm{r}_{\tilde{h}\tilde{h},\ell} \in \mathbb{C}^{N}$ the correlation vector. These statistics rely on the sample autocorrelation function  $\re{\varphi_{\tilde{h}\tilde{h}}[k] \approx \varphi_{h\re{h}}[k]}$, and they are defined as
\begin{subequations}
\begin{align}
\footnotesize\bm{R}_{\tilde{h}\tilde{h},\ell} = \re{\begin{bmatrix}  \varphi_{\tilde{h}\tilde{h}}[0] & \varphi_{\tilde{h}\tilde{h}}[1] & \cdots & \varphi_{\tilde{h}\tilde{h}}[N-1]\\
\varphi^{\ast}_{\tilde{h}\tilde{h}}[1] & \varphi_{\tilde{h}\tilde{h}}[0] & \cdots & \varphi_{\tilde{h}\tilde{h}}[N-2] \\
\varphi^{\ast}_{\tilde{h}\tilde{h}}[2] & \varphi^{\ast}_{\tilde{h}\tilde{h}}[1] &\cdots & \varphi_{\tilde{h}\tilde{h}}[N-3] \\ \vdots &  \vdots & \ddots & \vdots\\ \varphi^{\ast}_{\tilde{h}\tilde{h}}[N-1] & \varphi^{\ast}_{\tilde{h}\tilde{h}}[N-2] & \cdots & \varphi_{\tilde{h}\tilde{h}}[0] \end{bmatrix}}, \\
\smallskip
\bm{r}_{\tilde{h}\tilde{h},\ell} = \big[\re{\varphi_{\tilde{h}\tilde{h}}[\ell]\quad \varphi_{\tilde{h}\tilde{h}}[\ell+1]  \quad \cdots\quad \varphi_{\tilde{h}\tilde{h}}[N-1+\ell]}\big]^T.
\end{align}
\end{subequations}
We calculate the sample autocorrelation function for a lag $k$ using the $L_{\mathrm{train}}$ past LS estimates as
\begin{equation}
     \re{\varphi_{\tilde{h}\tilde{h}}[k]}=\re{\frac{1}{L_{\mathrm{train}}}}\sum_{n=0}^{L_{\mathrm{train}}-1}\tilde{h}[n]\tilde{h}^{\ast}[n+k].
\end{equation}

\begin{figure}[t]
    \centering
    \includegraphics[width=0.9\columnwidth]{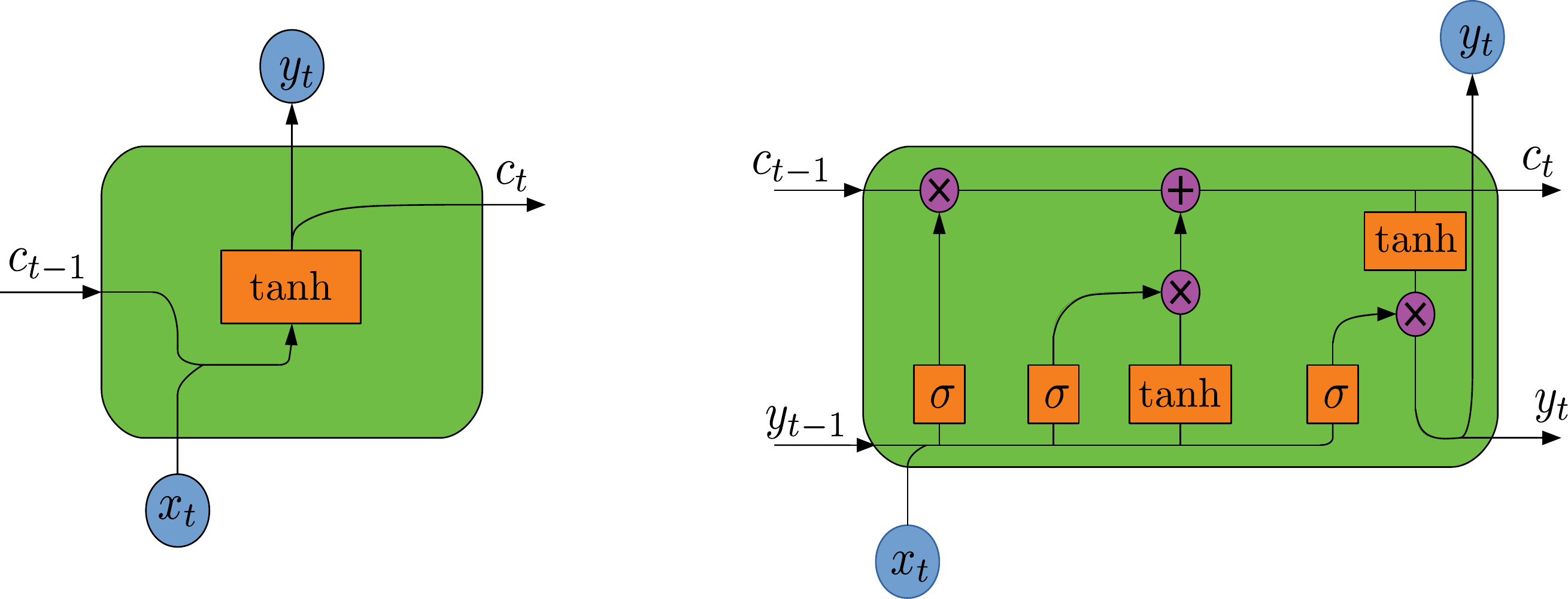}
    \caption{RNN (left) and LSTM (right) units with sigmoid ($\sigma$) and hyperbolic tangent activation functions. Note that they both take in the state $c_{t-1}$ from the previous unit and combine it with input $x_t$ to yield an output $y_t$ and a new state $c_t$, which is passed on to the next unit. The LSTM unit additionally takes in the previous input $y_{t-1}$.}
    \label{rnnlstm}
\end{figure}
\subsection{RNN and LSTM Predictors}

RNN and LSTM networks are a class of ML networks that contain feedback loops, allowing them to retain past information~\re{\cite{yu2019review}}. Due to their ability to maintain an internal state, they perform time series predictions with remarkable accuracy. They are typically trained using gradient descent with back-propagation, where a set of inputs is compared to the output based on a cost function and the prediction errors are propagated back through the network, leading to an update of weights and biases~\re{\cite{jiangNeuralNetworkBasedFading2019}}. This process is performed iteratively until the cost function converges to a minimum.

RNNs suffer from the so-called vanishing-gradient problem, whereby the gradient information that is back-propagated becomes too small such that it has little impact on the weights and biases. To solve this problem, LSTM units include additional ``gates'' which control the amount of information passing through the network \cite{hochreiterlstm}. Figure \ref{rnnlstm} illustrates the structure of RNN and LSTM unit\re{s}. 

\subsubsection{Training the Neural Networks}
In our channel prediction problem, the $L_{\mathrm{train}}$ complex-valued LS estimates are divided into \re{overlapping} blocks of $N$ samples to be fed into the neural networks during the training phase. Each complex-valued sample block is split into \re{its} real and imaginary part and fed into the network as $\re{2N}$ input. In particular, if $h_r[\cdot]$ and $h_i[\cdot]$ are the real part and the imaginary part of a channel sample, the input vector comprising $N$ samples is given by
\begin{align}
    \overline{\bm{h}}=\big[h_r[0],h_i[0],\cdots,h_r[N-1],h_i[N-1]\big]^T \in \mathbb{R}^{2N}.
\end{align}
Note that we use the same notation as that of the Wiener predictor \re{of} order $N$ to denote the number of input neurons $2N$. The neural network is then trained based on all $L_{\mathrm{train}}$ samples.

\subsubsection{Predicting the Channel Coefficients}
In the prediction stage, we feed the neural networks with $N$ complex-valued samples ($2N$ real-valued samples) from the $L_{\mathrm{test}}$ test samples, $N \leq L_{\mathrm{test}}$, to predict $h[n+\ell]$. The neural network yields a two-dimensional $\ell$-step prediction $\big[\hat{h}_r[n+\ell],\hat{h}_i[n+\ell]\big]^T$ at its output. The output is then re-synthesized into a complex-valued prediction as $\hat{h}[n+\ell] = \hat{h}_r[n+\ell]+\mathrm{j}\hat{h}_i[n+\ell]$. 


We emphasize that the considered ML-based predictors adjust their internal weights and biases directly from the LS estimates. By contrast, the Wiener predictor first needs to estimate the corresponding autocorrelation function, calculate \eqref{pred_eqn}, and finally predict the future channel state. Therefore, the neural networks are \emph{data driven}, while the Wiener predictor is based on statistics.



\section{Simulation Results}
\label{sec:simres}

In this section, we present performance comparisons between the Wiener filter and the ML-based predictors. We perform Monte Carlo (MC) simulations to evaluate the performance of the considered predictors. A single MC trial proceeds as follows:
\begin{enumerate}
    \item Generate a channel realization using the SOS method.
    \item Obtain $L$ noisy LS channel samples
    \item Design the predictors, i.e.,
    \begin{itemize}
        \item Train the neural networks using $L_{\mathrm{train}}$ samples.
        \item Estimate \re{the} autocorrelation function and calculate \re{the} Wiener predictor coefficients using $L_{\mathrm{train}}$ samples.
    \end{itemize}
    \item Predict \re{the} channel using $N$ estimates in the test set $\re{L_{\mathrm{test}}}$.
\end{enumerate}
We generate the flat-fading channel using the SOS method with $M=200$. The sampling period $T_s$ is set \re{to} 1 ms with a maximum Doppler spread $f_{\mathrm{D,max}}$ of 100 Hz. The ML-based predictors are trained using the Adam optimization algorithm \re{\cite{adam}} with a learning rate of 0.01, a hyperbolic tangent activation function, and the MSE as the loss function. Dropout regularization is added to reduce overfitting. We also use early stopping~\re{\cite{chollet2015keras}}, where the training process is halted if the loss function does not improve beyond a certain threshold. For simplicity and to save computation time, we limit our models to predict only one channel sample, as opposed to a sequence of samples. The reported results were obtained by averaging $100$ independent MC trials.

\begin{figure}[h]
\begin{minipage}[b]{0.5\columnwidth}
    \centering
    \includegraphics[width=0.95\linewidth]{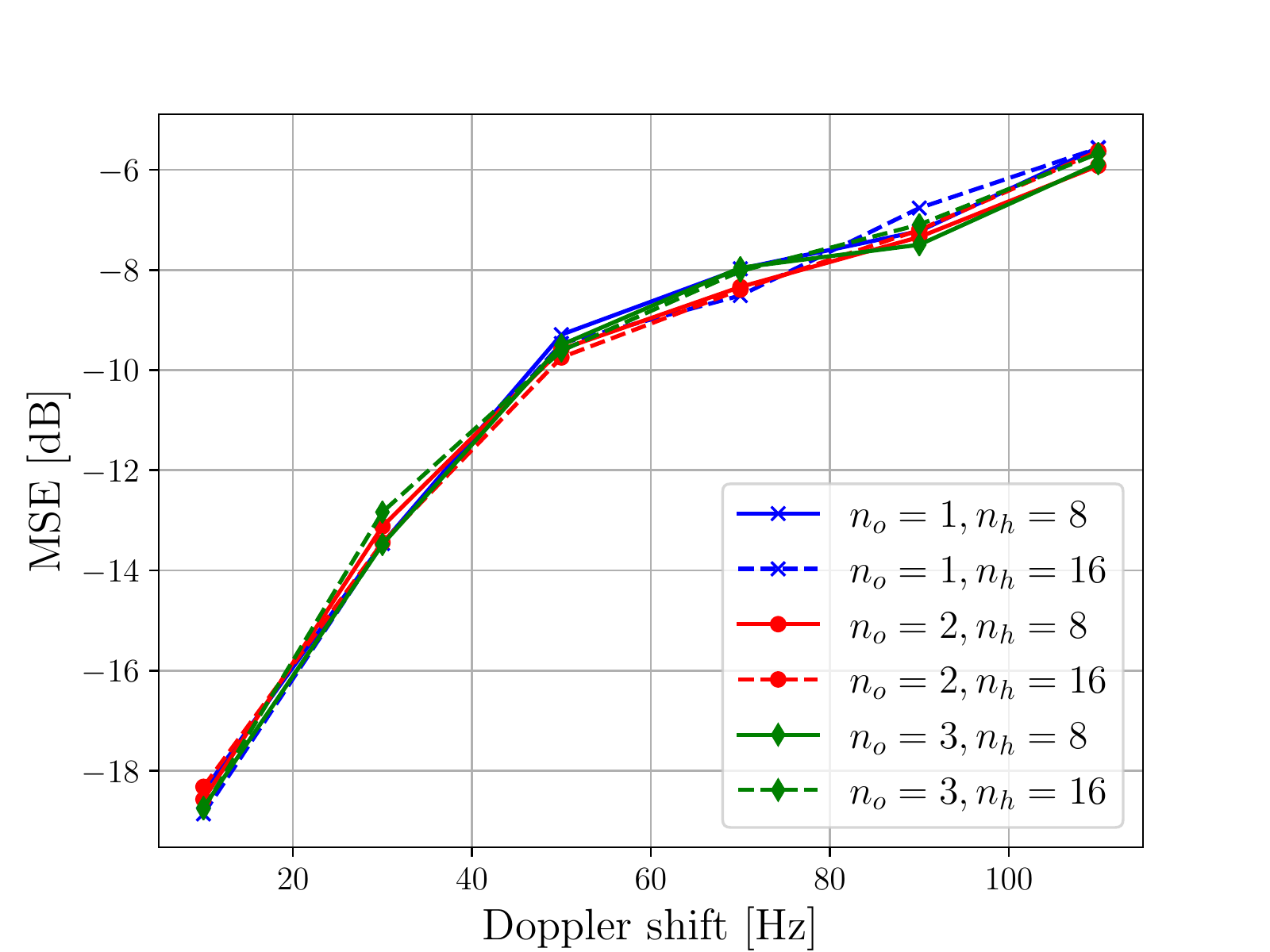}
    \subcaption{\label{1a}}
\end{minipage}
\hspace{-0.8em}
\begin{minipage}[b]{0.5\columnwidth}
    \centering
    \includegraphics[width=0.95\linewidth]{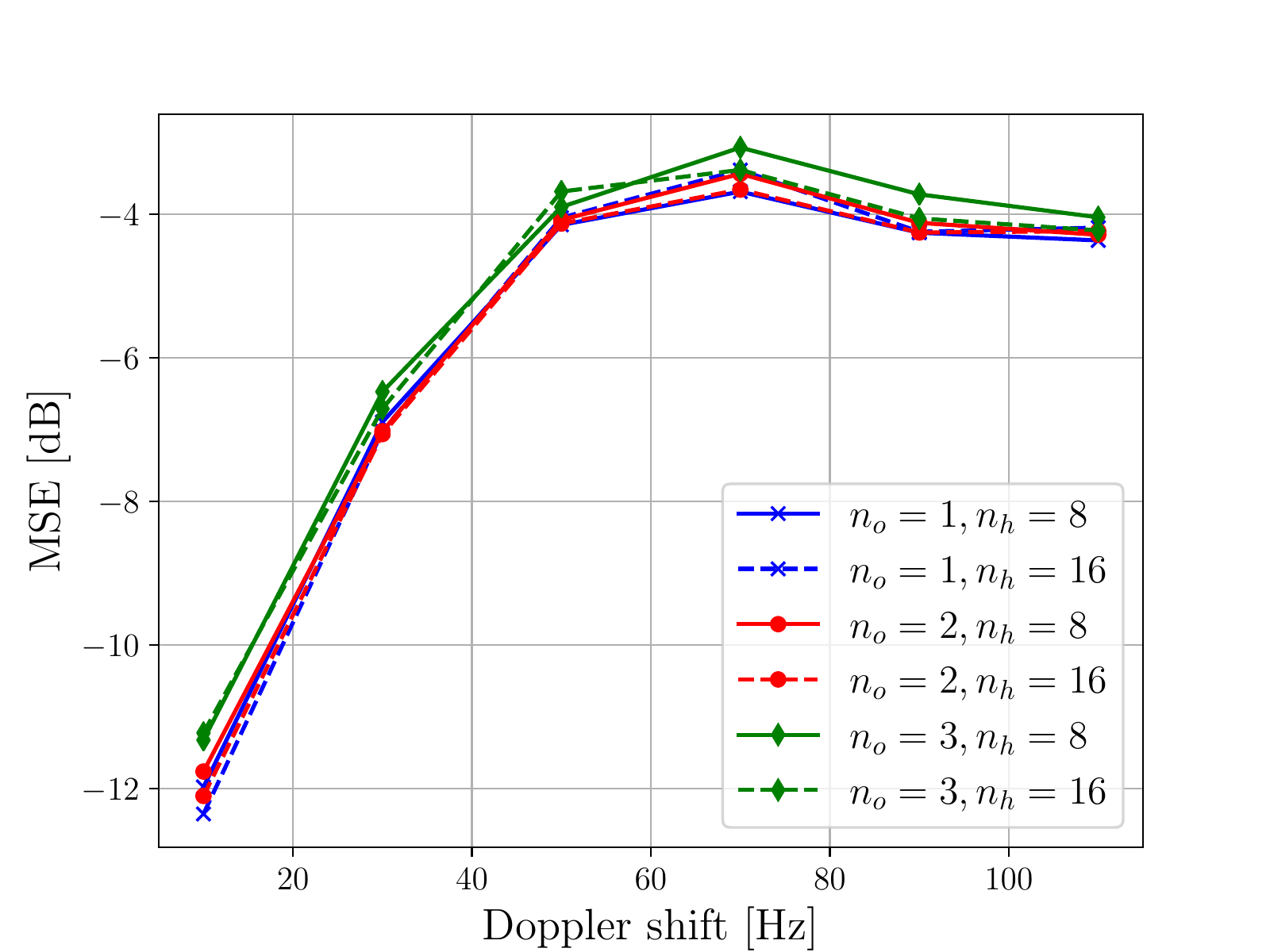}
    \subcaption{\label{1b}}
\end{minipage}
\caption{Model performance for various hidden layers $n_o$ and hidden neurons $n_h$ for: (a) $\ell=2$; and (b) $\ell=5$. SNR is set to 10 dB in both scenarios.}
\label{figs1}
\end{figure}

First, we evaluate the performance of six RNN predictors to gain insight into the number of \re{required} hidden layers $n_o$ and the number of \re{required} hidden neurons $n_h$. We obtain 500 LS estimates and use 75\% of them to train the models, while the rest of the samples are used for testing. Batch training is used with a batch size of 16 and the SNR is fixed at 10 dB. Figures \ref{figs1}\subref{1a} and \ref{figs1}\subref{1b} show the MSE performance of six models, with respect to Doppler shift, for prediction lengths $\ell = 2$ and $\ell=5$. Here, we fix the observation window length \re{to} $N=5$. All models perform roughly the same for $\ell = 2$, while for $\ell=5$, the models with $n_o = 1,2$ and $n_h = 16$ perform the best overall, especially for larger Doppler frequencies. Guided by this result, we set $n_o = 1$ and $n_h = 16$ for the rest of the simulations (for both the RNN and LSTM predictors), choosing one hidden layer instead of two to save on training time. These results indicate that simple models are \re{sufficient} to predict Rayleigh channels. We believe more complex models would be necessary to characterize other channel models, e.g., the 3GPP TR 38.901 channel model~\cite{3gpp2018study}.

\begin{figure}[h]
\begin{minipage}[b]{0.35\columnwidth}
    \centering
    \includegraphics[width=\linewidth]{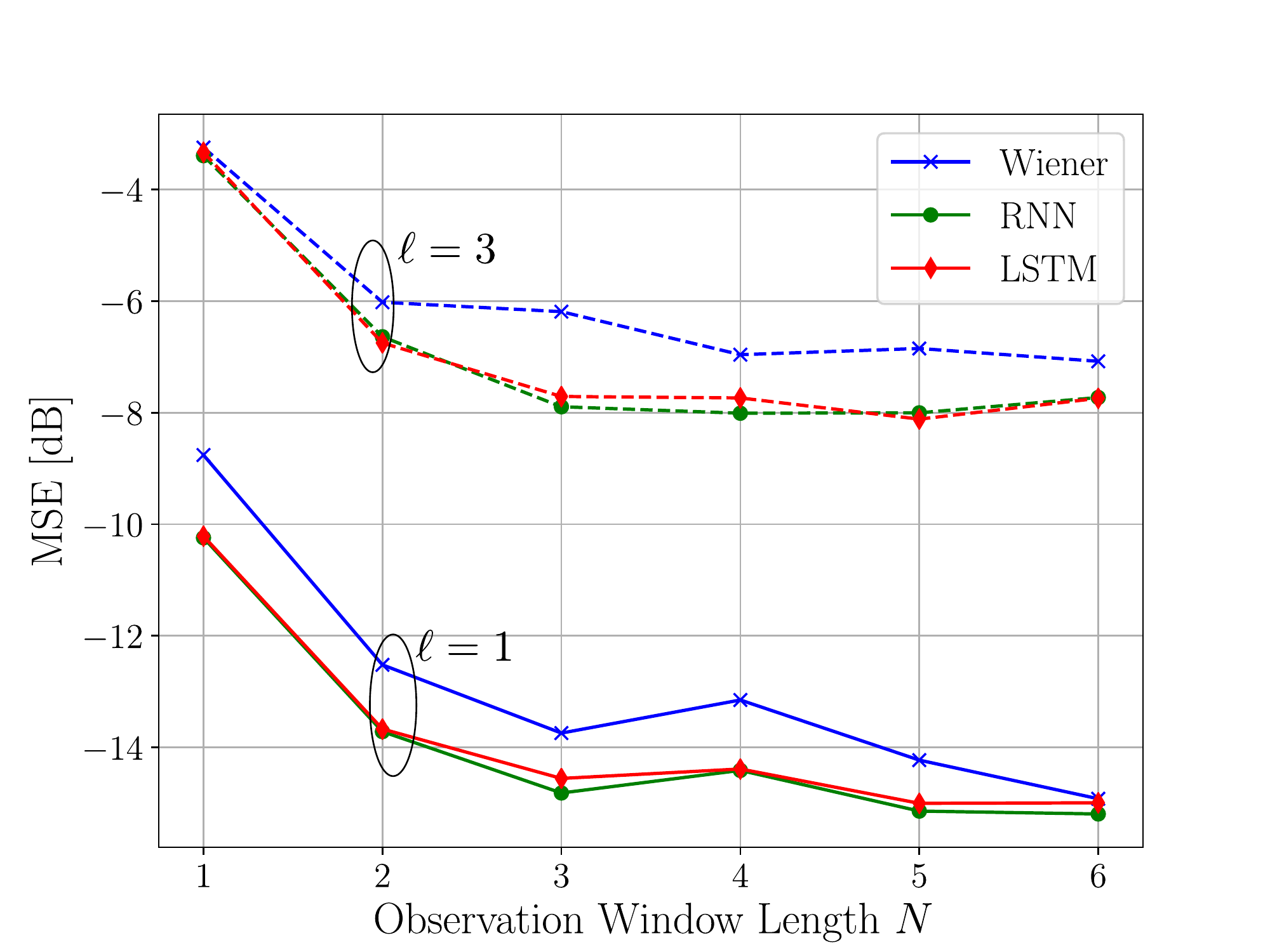}
    \subcaption{\label{2a}}
\end{minipage}
\hspace{-1.8em}
\begin{minipage}[b]{0.35\columnwidth}
    \centering
    \includegraphics[width=\linewidth]{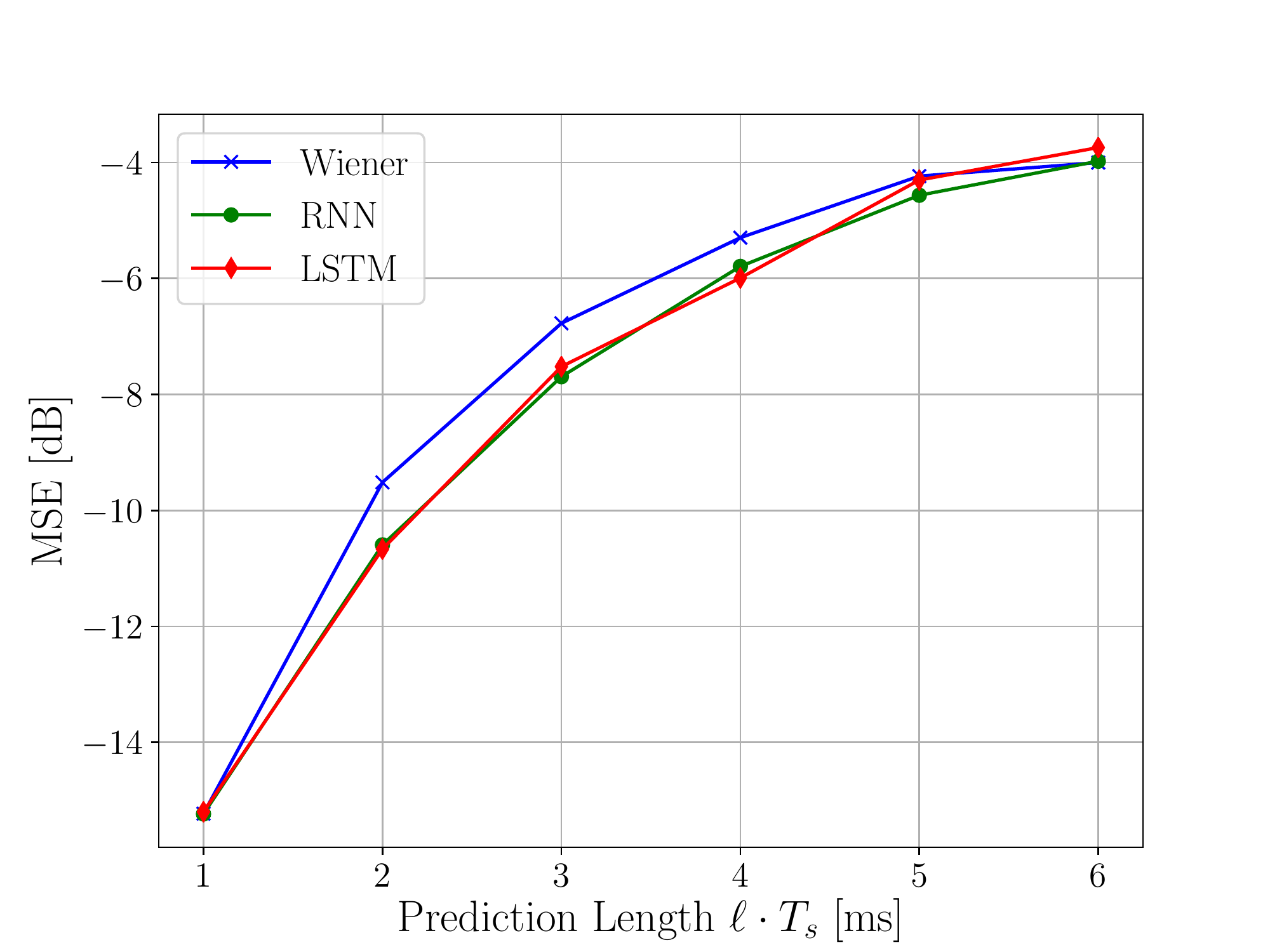}
    \subcaption{\label{2b}}
\end{minipage}
\hspace{-1.8em}
\begin{minipage}[b]{0.35\columnwidth}
    \centering
    \includegraphics[width=\linewidth]{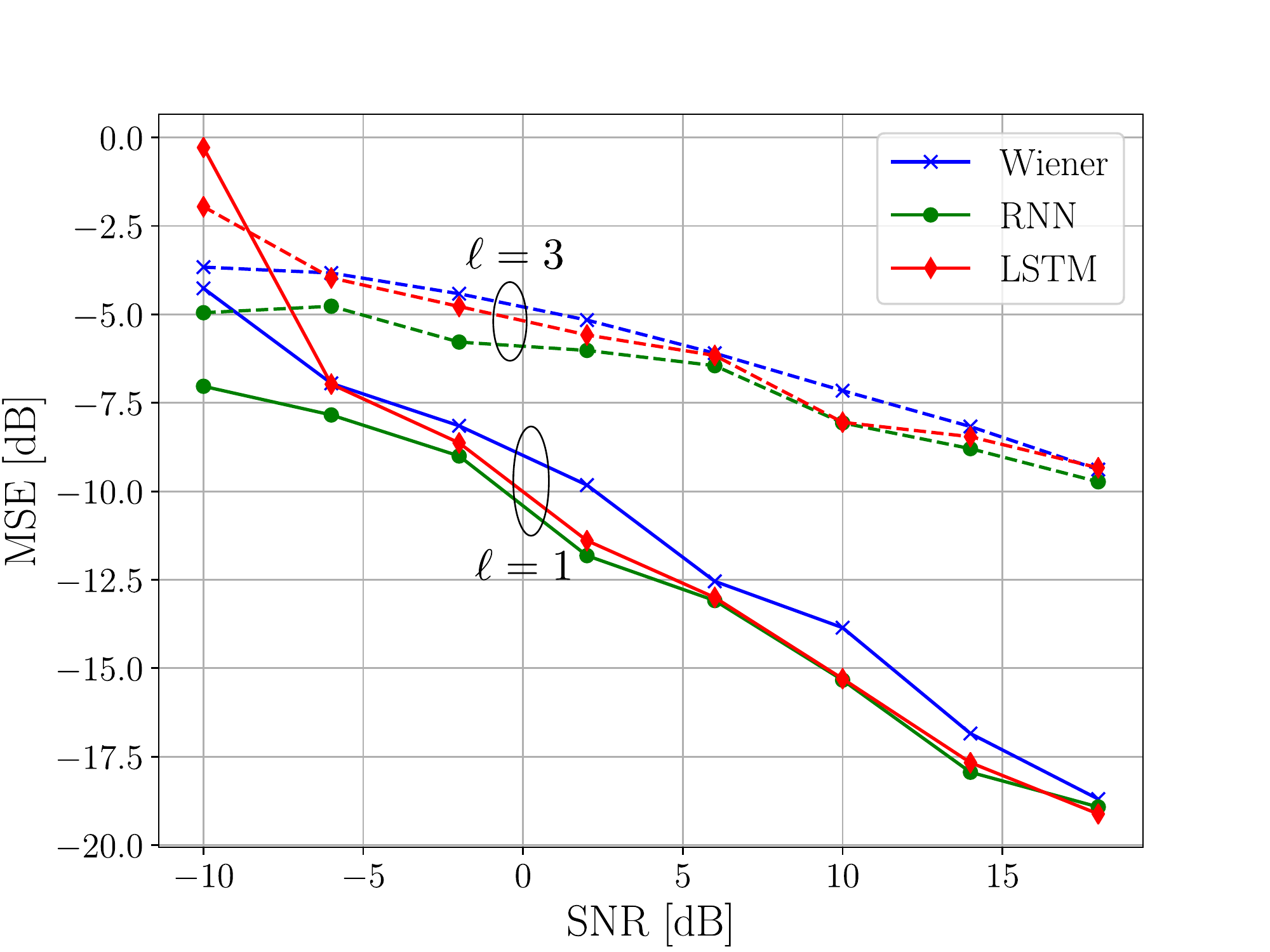}
    \subcaption{\label{2c}}
\end{minipage}
\caption{Performance comparisons with respect to: (a) observation window length (10 dB SNR); (b) prediction length (10 dB SNR, $N=5$); and (c) SNR ($N=5$).}
\label{figs2}
\end{figure}

In the second set of experiments, we evaluate the MSE for varying observation window length\re{s} $N$, prediction length\re{s} $\ell$ and SNR\re{s}. We obtain 1000 LS estimates and use 75\% for training the neural networks and \re{for} calculating the statistics \re{of} the Wiener predictor and the rest for testing. Figure \ref{figs2}\subref{2a} shows the MSE performance with respect to various observation window lengths $N$ for a fixed SNR of 10 dB. For $\ell = 1$, the ML-based predictors outperform the Wiener predictor for $N<5$, with the RNN predictor having a gain of about 1$~$dB compared to the Wiener predictor. The three predictors exhibit roughly the same performance at $N=6$. The ML-based predictors can learn the channel evolution even \re{from} a few observations while the Wiener predictor requires more observations to obtain accurate channel statistics. For $\ell=3$, the performance gain of the ML-based predictors is negligible. 
\onecolumn
%
\begin{multicols}{2}
\noindent


Figure \ref{figs2}\subref{2b} display the prediction MSE with respect to various prediction lengths $\ell$. The SNR is fixed at 10 dB and $N=5$. The gain provided by the ML-based predictors increases as $\ell$ increases and, for $\ell > 3$, the RNN predictor shows a considerable gain over the Wiener predictor. This reveals the robustness of ML-based predictors compared to the Wiener predictor even as the correlation between channel samples grows weaker with increasing $\ell$. The ML-based predictors perform better than the Wiener predictor over a wide range of SNRs, as shown in Figure \ref{figs2}\subref{2c} \re{(}here, $N=5$ as before\re{)}. The performance gain for the ML-based predictors is more pronounced for $\ell=1$ and decreases at high SNRs. Thus, the ML-based predictors are more robust in the presence of noise and provide more accurate predictions than the Wiener filter even at low SNRs. 

\section{Conclusions}
We \re{have} compared the prediction performance of the Wiener predictor and ML-based predictors in a Rayleigh flat-fading channel. We show that shallow RNN models are sufficient to characterize the channel over a wide \re{range of} Doppler shift\re{s} for the considered system model. Simulation results indicate that, for small prediction lengths, ML-based predictors outperform the Wiener predictor for small observation window lengths. They are also more robust under weak channel correlation and in the presence of noise, i.e., at large prediction lengths and low SNRs. These performance gains arise due to the non-linearities of the neural networks and their data-driven nature, as opposed to the Wiener predictor, which is linear and dependent on statistical knowledge of the channel. \re{However, for small observation window lengths with a fixed large prediction length, as well as for high SNRs, the ML-based predictors and the Wiener predictor perform roughly the same.}

Our results indicate that ML-based channel prediction is a promising solution that outperforms the classical Wiener filter in some scenarios. In future work, we intend to investigate this approach in practical systems, e.g., 5G NR, and intelligent reflective surface-assisted systems.

%
\bibliographystyle{IEEEtran}
\bibliography{myRefs}
\end{multicols}
\end{document}